\documentclass[aps,prb,twocolumn,amsmath]{revtex4}
\usepackage{graphicx}
\usepackage{amssymb}
\usepackage{bm}
\usepackage{kotex}

\begin{document}
\setcounter{page}{1}

\title{Entanglement Properties of the One-Dimensional Dimerized Fermi-Hubbard Model}
\author{ Min-Chul Cha$^{1}$\footnote{corresponding author: mccha@hanyang.ac.kr}, 
Hoon Beom Kwon$^{1}$, Ji-Woo Lee$^{2}$, Myung-Hoon Chung$^{3}$}
\affiliation{ $^{1}$Department of Photonics and Nanoelectronics, Hanyang University, Ansan 15588, Korea\\ 
$^{2}$ Department of Physics, Myongji  University, Yongin 17058, Korea\\
$^{3}$ College of Science and Technology, Hongik  University, Sejong 30016, Korea}
\begin{abstract}
We study the entanglement properties of the one-dimensional dimerized Fermi-Hubbard model. 
Using a matrix product state approach, we compute the ground state and identify two insulating phases at 1/2- and 3/4-filling, 
along with a metallic phase. 
The underlying physical mechanisms of these phases are conclusively characterized by their entanglement spectra. 
Our findings demonstrate that the two insulating phases are physically distinct: 
the phase at 1/2-filling possesses a charge gap originating from the robust band gap enhanced by repulsive interactions, 
whereas the phase at 3/4-filling exhibits a Mott gap resulting from electron interactions. 
This fundamental distinction is faithfully reflected in the finite-entanglement scaling properties of the half-chain entanglement entropy 
and the universal distribution of the entanglement spectrum.
\end{abstract}
\keywords{quantum phase transition, entanglement, matrix product states}
\maketitle

\section{Introduction}
Entanglement properties provide a powerful framework for exploring and understanding interacting quantum systems
\cite{Amico08,Horodecki09,Calabrese09R,Laflorencie16}.
As one of the most distinctive features of quantum systems, entanglement marks quantum phase transitions
and offers useful tools for analyzing the critical properties through the universal scaling behavior of entanglement entropy (EE)
\cite{Calabrese09,Korepin04,Pollmann09,Tagliacozzo08}
and the universal distribution of the entanglement spectrum (ES)
\cite{Calabrese08,Alba18,Cha22}.
It is also useful for characterizing quantum phases, including topological properties
\cite{Kitaev06,Levin06},
which are often closely related to the energy gaps and the edge states that these gaps entail
\cite{Hasan10,Qi11,Bansil16,Qi12,Sanpera12}.
Significant efforts have been devoted to understanding how interactions modify these properties
\cite{Rachel18,Tang12,Manmana12,Guo11,Ara24,Wang15,Wagner23,Phillips23}.

The dimerized Fermi-Hubbard model, 
originally introduced to study the physical properties induced by dimerization in materials such as polyacetylene in the non-interacting limit
\cite{Su79,Asboth16},
serves as one of the simplest models for testing the usefulness of entanglement properties in characterizing quantum phases of interacting systems. 
This model is notable for exhibiting two different insulating phases arising from the interplay of interaction and dimerization, 
each with distinct properties, including topological aspects
\cite{Le20,Mikhail24}.
Entanglement properties provide a comprehensive approach for revealing the fundamental differences between these two insulating phases. Consequently, many studies have focused on clarifying the bulk-edge correspondence through the EE in this model
\cite{Ryu06,Sirker14,Bisht24,Zhou25,Chang25}, 
and entanglement spectra have also been employed to study the topological properties~\cite{Ye16}.

In this work, we study the dimerized Fermi-Hubbard model in one dimension, employing a matrix product state (MPS) approach
\cite{Vidal04,Vidal07,Cha18} to calculate the ground state.
The ground state exhibits the two distinct insulating phases, at 1/2-filling and at 3/4- (or 1/4-) filling.
At 1/2-filling, the charge gap arises from the band gap in the weak interaction region, which is further enhanced by repulsive on-site interactions.
(In the strong interaction limit, a Mott gap opens with the presence of antiferromagnetic ordering.
Here, we focus on the case of moderately weak interactions.)
On the other hand, at 3/4-filling, the Mott gap primarily results from interactions among particles.
The differences between these two insulating phases elucidate the underlying physics of strongly correlated systems.

To investigate the differences, we analyze the entanglement properties of the system. 
The EE serves as a practical macroscopic probe for locating the phase transition lines and constructing the phase diagram.
Our findings suggest that arbitrarily small repulsive interactions can open a charge gap in the 3/4-filled insulating phase.
The scaling behavior of the EE distinctly clarifies the different nature of these two insulating phases.
Furthermore, while the EE provides an intuitive benchmark for the edge modes in the fully dimerized limits, 
the definitive identification of the topological and physical features of these phases is provided by the ES.
In addition, each phase, including the metallic phase, can be identified by its characteristic structure appearing in the ES.
The universal scaling and distribution of the ES further highlight the different physical nature of the two insulating phases.

\section{One-dimensional dimerized Fermi-Hubbard model}
Let us consider the Hamiltonian of the one-dimensional dimerized Fermi-Hubbard model with staggered hopping amplitudes
$t_+=t_0+\delta t$ and $t_-=t_0-\delta t$, given by
\begin{eqnarray}
H=&-&\sum_{ij,\sigma}{\bold c}_{i\sigma}^\dagger\cdot {\bold T}_{ij}\cdot {\bold c}_{j\sigma}\cr
&+&U\sum_i (n_{iA,\uparrow}n_{iA,\downarrow}+ n_{iB,\uparrow}n_{iB,\downarrow})
\label{eq:H}
\end{eqnarray}
with
\begin{eqnarray}
{\bold T}_{ij}=
\left( \begin{array}{cc}
(\mu+\frac{U}{2})\delta_{ij} &  t_+\delta_{ij}+t_-\delta_{i,j+1}\\
t_+\delta_{ij}+t_-\delta_{i,j-1} & (\mu+\frac{U}{2})\delta_{ij}\end{array}\right)
\end{eqnarray}
and 
\begin{eqnarray}
{\bold c}_{j\sigma}=
\left( \begin{array}{c}
c_{jA,\sigma}\\
c_{jB,\sigma}\end{array}\right),
\end{eqnarray}
where we take two adjacent sites (sites $A$ and $B$, belonging to $A$- and $B$-sublattices, respectively) coupled  via $t_+$ as a unit cell.
Here, $i$ and $j$ are indices for the  unit cells.
The operators $c_{jA,\sigma}(c_{jB,\sigma})$ and $c_{jA,\sigma}^\dagger (c_{jB,\sigma}^\dagger)$ 
represent the particle annihilation and creation operators, respectively, and 
$n_{jA,\sigma}(n_{jB,\sigma})$ denote the particle number operators at sites $A$ and $B$ in the $j$-th cell,
with $\sigma=\uparrow,\downarrow$ representing the spin index.
$U$ is the on-site repulsion strength, and $\mu$ is the chemical potential.
Here, the chemical potential is shifted by $U/2$ so that $\mu = 0$ corresponds to the particle-hole symmetry line. 
This conventional choice explicitly reflects the symmetry of the system, resulting in a phase diagram that is symmetric with respect to $\mu = 0$.
We consider $t_+$ as the intra-cell coupling, while $t_-$ serves as the intercell coupling.
For convenience, we set $t_0=1$ in our study.

\subsection{$U=0$ limit}
In the non-interacting limit ($U=0$), known as the Su-Schrieffer-Heeger model, 
the Hamiltonian can be diagonalized using Fourier transforms
\begin{eqnarray}
{\bold {\tilde c}}_{k\sigma}=\frac{1}{\sqrt{L/2}}\sum_j e^{i k R_j}{\bold c}_{j\sigma},\cr
{\bold c}_{j\sigma}=\frac{1}{\sqrt{L/2}}\sum_k e^{-i k R_j}{\bold {\tilde c}}_{k\sigma},
\label{eq:ck}
\end{eqnarray}
which yields the following form:
\begin{eqnarray}
H_0=-\sum_{k,\sigma}{\bold {\tilde c}}_{k\sigma}^\dagger\cdot 
\left( \begin{array}{cc}
0 &  \tau_k e^{i\theta_k}\\
\tau_k e^{-i\theta_k}& 0\end{array}\right)
\cdot {\bold {\tilde c}}_{k\sigma},
\end{eqnarray}
where
\begin{eqnarray}
\tau_k=2 [\cos^2k+|\delta t|^2 \sin^2k]^{1/2},
\end{eqnarray}
and $\theta_k=\tan^{-1}[t_-\sin 2k /( t_++t_-\cos 2k)]$.
Here, $L$ is the number of sites, $R_j=2j$ denotes the position of the $j$-th unit celll for $j=1,\dots,L/2$, 
and $k=2\pi n/ L$ for an integer $n$ ($ - L/4 < n \le L/4$) is the wave number.

Under the periodic boundary condition with an even integer $L$, 
which is a prerequisite to preserve the bipartite nature of the lattice, 
the energy levels are given by
\begin{eqnarray}
\epsilon_\pm (k)=\pm \tau_k.
\end{eqnarray}
A finite $\delta t$ thus yields a band gap $\Delta \epsilon = 4| \delta t|$ at the Brillouine zone boundary ($k= \pi/2$),
resulting in a band insulator at $1/2$-filling, which is a typical example of a topological insulator~\cite{Asboth16}.
Even for $U >0$, we still expect that this band insulating property remains at $1/2$-filling,
with an enhanced charge gap due to the repulsive interactions for sufficiently large $|\delta t|$.

\subsection{$U>0$}
The most significant change introduced by $U>0$ is the opening of the Mott gap at 3/4- (or 1/4-) filling when there is a finite $|\delta t|$.
To investigate this insulating phase, we employ the MPS representation to effectively handle the interactions.
This method also provides entanglement properties that are very useful for identifying the different phases.

A canonical form of the MPS wavefunction~\cite{Vidal04,Vidal07} for the ground state is constructed as follows:
\begin{eqnarray}
|\Psi\rangle =\sum_{\{s_i\}}
&&A^{[s_1]}\Lambda_1 B^{[s_2]}\Lambda_2 C^{[s_3]}\Lambda_1 D^{[s_4]}\Lambda_2 A^{[s_5]}
\cdots D^{[s_L]}\Lambda_2\cr
&&\times|s_1,s_2,s_3,s_4,\cdots,s_L\rangle,
\label{eq:MPS}
\end{eqnarray}
where the matrices $A$, $B$, $C$, and $D$ are  $\chi \times \chi$ MPS matrices, 
$\Lambda_1$ and $\Lambda_2$ are diagonal $\chi \times \chi$ matrices (where $\Lambda_{\alpha\beta}=\lambda_{\alpha}\delta_{\alpha\beta}$ for each $\Lambda$),
and $s_\ell$ denotes physical index for the basis states 
($|s_\ell\rangle =|0\rangle,|\uparrow\rangle,|\downarrow\rangle,|\uparrow\downarrow\rangle$) at the $\ell$-th site.
We take the limit $L\to \infty$ to consider an infinite chain.
To obtain the correct wavefunction, we need to repeatedly multiply matrices 
in the form $A^{[s_a]}\Lambda_1 B^{[s_b]}\Lambda_2 C^{[s_c]}\Lambda_1 D^{[s_d]}\Lambda_2$
especially near or at 3/4-filling.
This differs from the form $A^{[s_a]}\Lambda B^{[s_b]}\Lambda$ used when $\delta t=0$.
If we attempt to use the form $A^{[s_a]}\Lambda_1 B^{[s_b]}\Lambda_2 C^{[s_c]}\Lambda_3 D^{[s_d]}\Lambda_4$
with four different $\Lambda$'s, 
we consistently find that $\Lambda_1=\Lambda_3$ and $\Lambda_2=\Lambda_4$.
We adopt a periodic boundary condition and take the limit where the system size $L \to \infty$
to construct an MPS ansatz for an infinite chain.

The lowest energy state is then obtained by using the time-evolving block decimation (TEBD) algorithm~\cite{Vidal04,Vidal07}.
By dividing the imaginary time into small intervals of size $\Delta\tau$,
we apply the the Suzuki-Trotter decomposition $e^{-\tau H}=(e^{-\Delta\tau H})^N (\Delta\tau=\tau/N)$
to implement the two-site TEBD~\cite{Cha22}.
The number of repetitions $N$ is determined under the condition that the energy difference between subsequent states is less than 
$10^{-12}$.
A smaller $\Delta\tau$ reduces the decomposition errors but requires a longer time for the TEBD process.
We choose $\Delta\tau=0.001$ in our calculations, which costs a long time to compute
but is necessary to obtain the lowest energy state, especially near the phase transition.

The diagonal elements of the matrix $\Lambda$ represent the Schmidt coefficients, which are non-negative real numbers.
The half-chain EE, $S_h$, is then defined as
\begin{eqnarray}
S_h= -\sum_{\alpha=1} \lambda_\alpha^2 \log_2 \lambda_\alpha^2,
\label{eq:EE}
\end{eqnarray}
where $\lambda_\alpha$ is the $\alpha$-th diagonal element of $\Lambda$ which bipartites the one-dimensional system.
We also construct the reduced density matrices for single-site and two-site subsystems,
and diagonalize them to obtain the corresponding EEs.

\section{Entanglement Properties}
\begin{figure}[t] 
\includegraphics[width=0.45\textwidth]{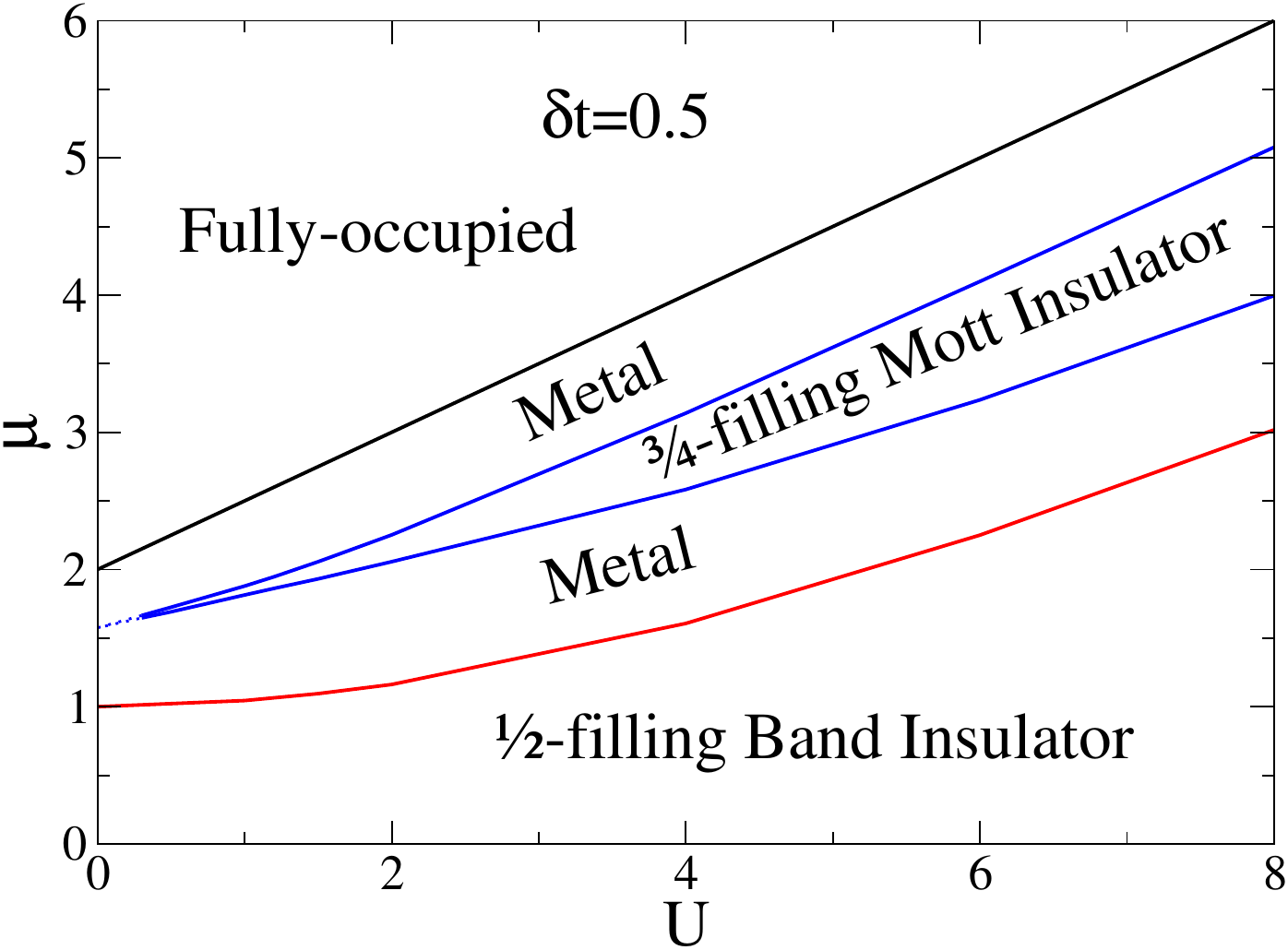}
\caption{Schematic phase diagram of the one-dimensional dimerized Fermi-Hubbard model at a dimerization strength of $\delta t=0.5$.
At 1/2-filling, a band insulating phase with a band gap $\Delta=4\delta t$ emerges when $U=0$, 
whereas a Mott insulating phase appears at 3/4- (or 1/4-) filling.}
\label{fig:1dFHM_PD}
\end{figure}

The ground-state wavefunctions obtained through MPS calculations with a bond dimension of  $\chi=40$ yield the EEs and density, 
from which we can construct a phase diagram. 
Figure~\ref{fig:1dFHM_PD} shows a schematic phase diagram of the model for $\delta t=0.5$.
In the metallic phase, the density varies continuously as a function of the chemical potential $\mu$. 
In contrast, in the insulating phase, the density remains constant even as $\mu$ varies.
In addition to the insulating phase at 1/2-filling, another insulating phase appears at 3/4-filling,
which is characteristic of the dimerized Fermi-Hubbard model.
Within our numerical accuracy, the gap opens for an arbitrarily small $U$.
The charge gap at 1/2-filling is a band gap resulting from dimerization, in contrast to the ordinary Fermi-Hubbard model
\cite{Korepin05Book} (i.e., $\delta t=0$), 
where the charge gap is a Mott gap arising from interactions. 
On the other hand, at 3/4-filling, the Mott phase emerges due to interactions in the presence of dimerization.

\begin{figure}[t] 
\includegraphics[width=0.45\textwidth]{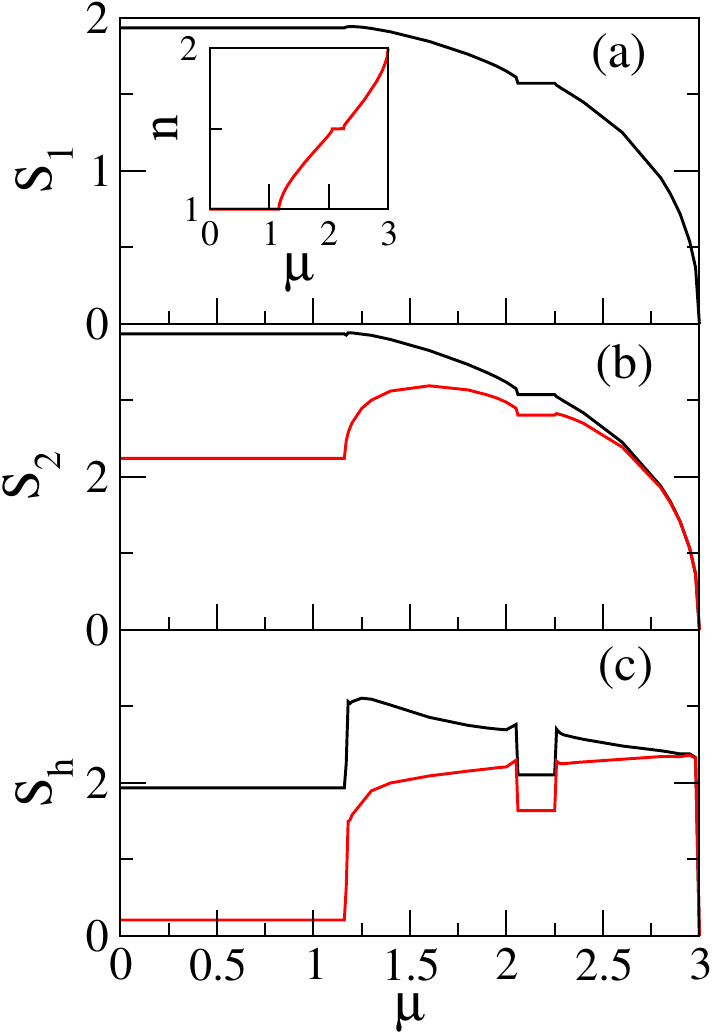}
\caption{EEs for (a) single-site, (b) two-site, and (c) half-chain subsystems as a function of $\mu$ at $U=2$ and $\delta t=0.5$. 
The calculation is performed with a bond dimension of $\chi=40$. 
The black and red lines represent the EE when the subsystem is coupled to the rest of the system via a $t_{+}$-bond and a $t_{-}$-bond, respectively; 
in (a), these lines overlap. 
The inset illustrates the density behavior, clearly identifying the transitions between the insulating phase at fixed density and the metallic phase.}
\label{fig:1dFHM_S}
\end{figure}

Figure~\ref{fig:1dFHM_S} shows the typical EEs for how much (a) a single-site, (b) a two-site, or (c) a half-chain subsytem
is correlated with the rest of the lattice when $U=2$ and $\delta t=0.5$.
The upper (black) lines represent the EE when the subsytem is coupled to the rest of the system via a $t_+$-bond, 
while the lower (red) lines correspond to the case of coupling via a $t_-$-bond.
A single site always has one $t_+$-bond and one $t_-$-bond, ensuring that the black and red lines overlap.
This behavior of EEs, along with the density behavior depicted in the inset,
allows us to easily identify the occurrence of phase transitions.
We can observe small peaks near the transition points, 
which are simply a consequence of the cluster mean-field behavior~\cite{Cha22} that occurs in MPS numerical calculations with a finite bond dimension.

At 1/2-filling, we specifically observe that $S_h \lesssim 2$ for $t_+$-bond and $S_h \gtrsim 0$ for $t_-$-bond,
which approach $2\log_2 2$ and 0, respectively, as $U \to 0$ in the limit of $|\delta t| \to 1$.
It is crucial to emphasize that these specific values strictly correspond to the idealized fixed-point limit, 
reflecting the localized degrees of freedom of the protected zero-energy edge states or the fully trivial state, respectively~\cite{Ryu06,Sirker14}.
Therefore, our characterization of the topological phase does not rely merely on the difference between the two EEs, 
but on its systematic evolution from these non-trivial fixed-point values.

Although strong interactions can induce a drastic structural change in these edge modes~\cite{Bisht24},
our results demonstrate that in the moderately interacting regime, 
the EE expectedly deviates from these specific values due to interaction-induced corrections. 
However, the definitive evidence for the survival of the protected edge states is provided by the full ES discussed below, 
which exhibits a persistent double degeneracy for the $t_{+}$-bonds even for finite $U$. 
Since the EE is a collective manifestation of these underlying eigenvalues,
its smooth and non-vanishing behavior—anchored by the unbroken ES degeneracy—provides a resilient signature 
that the topological boundary features remain robust against moderate on-site Coulomb repulsions.

This protection is consistent with the robust nature of the interacting symmetry-protected topological (SPT) phases in 1D systems,
where topological edge modes remain stable against moderate on-site Coulomb interactions~\cite{Manmana12,Le20,Chang25,Anfuso07,Barbiero18}. 
Similarly, the two-site EE $S_2$ approaches $4$ and $2$ in the fully dimerized limit, respectively, 
reflecting the localized dimeric nature of the ground state 
and further supporting the stable behavior of the subsystem entanglement properties under finite interactions~\cite{Peschel09,Cho17}.

However, these features are absent at 3/4-filling.
At this filling, the alternating hopping amplitudes of the non-interacting SSH model fail to open a single-particle gap at the Fermi level, 
stabilizing a gapless metallic phase instead. 
When a finite on-site interaction $U > 0$ is introduced, the system enters a Mott insulating phase characterized by a finite charge gap
but a remaining gapless spin sector, as extensively investigated in recent studies 
on correlated fractional fillings~\cite{Wang25, Mikhail24}.
Due to these remaining gapless spin degrees of freedom, the true half-chain EE (without MPS truncation) 
logarithmically diverges in the thermodynamic limit. 

As mentioned above, we can identify two distinct insulating phases where the EE and density remain unchanged as $\mu$ varies.
This implies that the wavefunctions are fixed in each insulating phase.
It raises an intriguing question of whether these two insulating phases are fundamentally different.
To clarify this, we investigate their entanglement properties.
The behavior of a diverging correlation length $\xi$ can be captured by the scaling relation
$S_h \sim (c/6)\log_2 \xi$ in one-dimensional systems, where $c=1$ is the central charge
\cite{Calabrese09,Korepin04}.
Furthermore, in an MPS with a finite bond dimension \(\chi \), 
which determines the size of the matrix used,
\(\xi \) can be characterized by \(\xi \sim \chi^\kappa\) with an exponent \(\kappa \) ~\cite{Pollmann09,Tagliacozzo08}.

\begin{figure}[t] 
\includegraphics[width=0.45\textwidth]{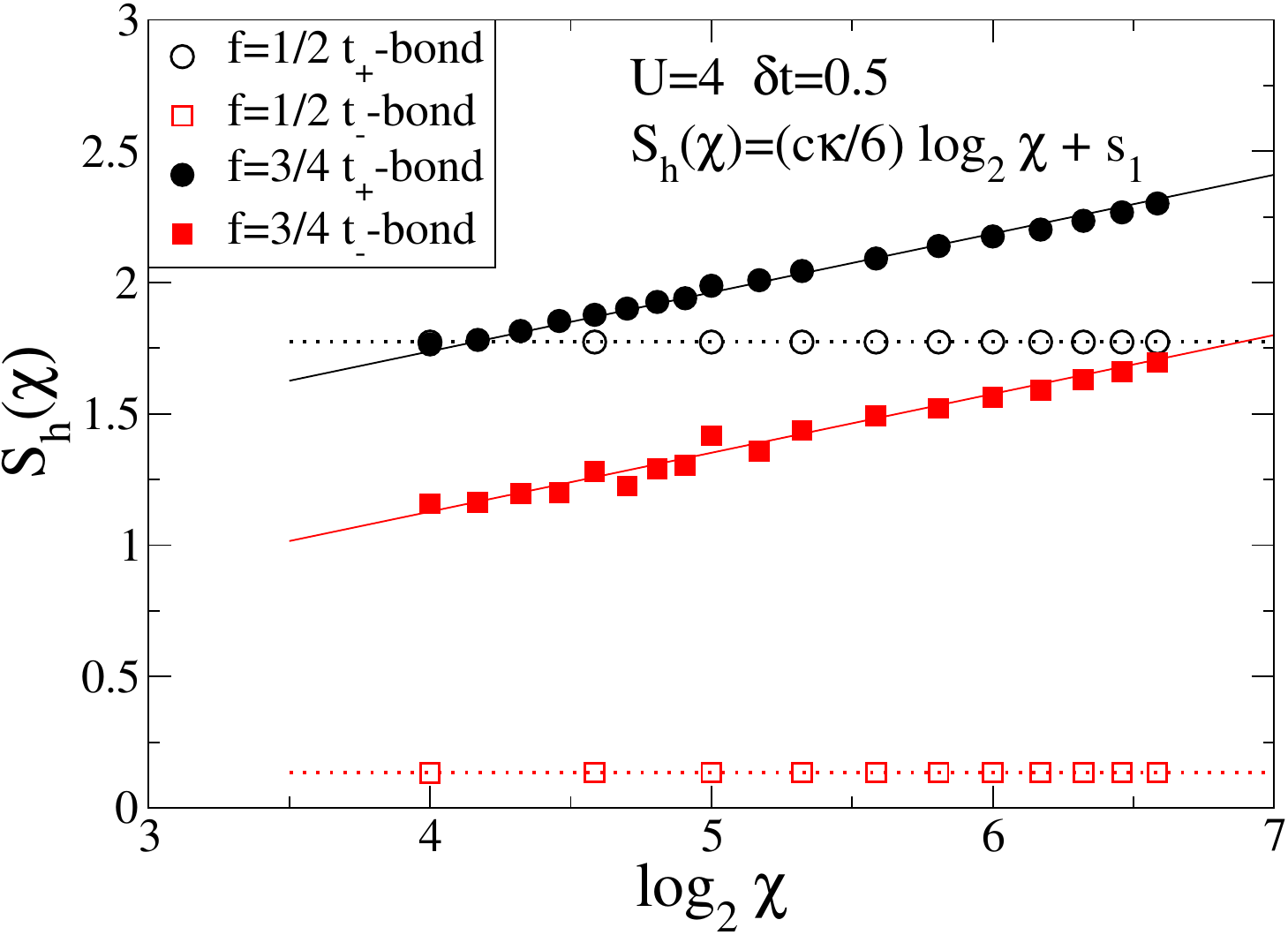}
\caption{Half-chain EE $S_h(\chi)$ in the insulating phases
as a function of the bond dimension $\chi$ for $U=4$ and $\delta t=0.5$.
At 3/4-filling, $S_h(\chi)$ exhibits a logarithmic scaling behavior indicative of a Mott insulating phase,
which is characterized by a correlation length scaling with the exponent $\kappa=1.344$. 
In contrast, at 1/2-filling, $S_h(\chi)$ is nearly independent of $\chi$,
reflecting the nature of a gapped band insulator with a finite correlation length. 
This contrast highlights the distinct physical nature of the two insulating phases.}
\label{fig:1dFHM_Shchi}
\end{figure}

Figure~\ref{fig:1dFHM_Shchi} illustrates the half-chain EE as a function of the bond dimension $\chi$ in the insulating phases.
The behavior at 3/4-filling follows the relation $S_h(\chi)=(c\kappa/6)\log_2 \chi +s_1$ with $\kappa=1.344$ (determined by the theoretical prediction~\cite{Pollmann09}),
where $s_1$ is a constant.
This logarithmic scaling behavior provides clear evidence of a Mott insulating phase characterized by a diverging correlation length,
reminiscent of the Mott phase observed in the ordinary half-filled Fermi-Hubbard model~\cite{Cha18}.
Conversely, at 1/2-filling, $S_h(\chi)$ is found to be almost independent of $\chi$,
indicating that the correlation length in this insulating phase remains short and finite, which is a typical signature of a gapped band insulator.
Thus, the scaling behavior of the half-chain EE distinctly demarcates the fundamental difference between the two insulating phases.

\begin{figure}[t] 
\includegraphics[width=0.45\textwidth]{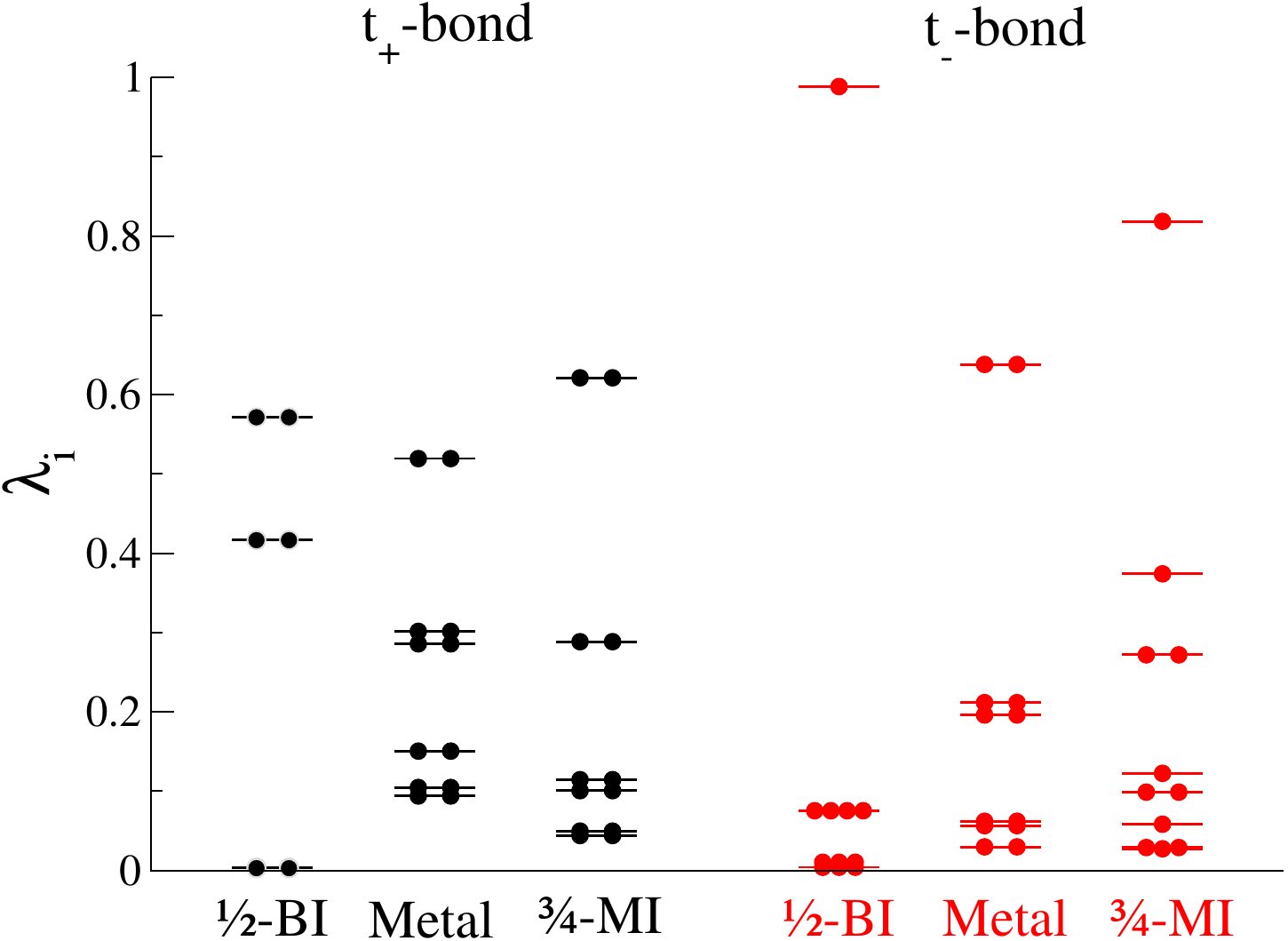}
\caption{Typical half-chain ES for $t_+$- and $t_-$-bonds in different phases of the dimerized Fermi-Hubbard model
at $U=2$ and $\delta t=0.5$.
In the metallic phase (at $\mu=1.6$), the ES exhibits a double degeneracy for both bonds.
This degeneracy is lifted in the insulating phase for $t_-$-bonds,
whereas $t_+$-bonds retain their double degeneracy, reflecting the spin-1/2 isospin state at 3/4-filling and the protected spin-1/2 edge states at 1/2-filling, respectively.}
\label{fig:1dFHM_Ses}
\end{figure}

The entanglement properties of the phases are clearly revealed in the ES,
which represents the distribution of eigenvalues of the reduced density matrix.
Figure~\ref{fig:1dFHM_Ses} displays the typical half-chain ES in different phases for $t_+$- and $t_-$-bonds.
Let us discuss the characteristic features of the ES in each phase:

(1) We observe that the ES in the metallic phase away from 1/2-filling exhibits a double degeneracy for both $t_+$- and $t_-$-bonds.
This double degeneracy is a hallmark of the metallic phase,  consistent with observations in the ordinary Fermi-Hubbard model \cite{Cha18}.
It is important to note that this degeneracy in the conducting regime is non-topological in nature. 
Rather, as suggested by recent theoretical frameworks for one-dimensional gapless phases \cite{Verresen17, Xu26}, 
such behavior can be largely attributed to robust projective symmetry constraints that persist even after the closure of the bulk charge gap. 
This implies that the observed double degeneracy in the metallic phase arises from general symmetry constraints in spinful chains
rather than signifying a topologically non-trivial insulating phase.

(2) This double degeneracy is lifted in the insulating phases for  $t_-$-bonds, as also occurs in the ordinary Fermi-Hubbard model \cite{Cha18},
while the ES for $t_+$-bonds still maintains a double degeneracy.
Notably, at 1/2-filling, this remaining double degeneracy provides compelling evidence
for a 1D symmetry-protected topological (SPT) phase, distinguishing it from a trivial insulating state.
Such degeneracies are widely recognized as a definitive signature of SPT order, as robustly established via the MPS framework ~\cite{Pollmann10,Pollmann12},
which is intimately linked to edge-state fractionalization in fermionic systems~\cite{Kitaev11,Pollmann11}. 
For Hubbard-type models, the classification of these SPT orders has been extensively demonstrated across various models~\cite{Verresen17,Montorsi17}. 
Accordingly, the double degeneracy observed  in the ES of our dimerized Fermi-Hubbard model confirms
the non-trivial topological robustness of the half-filled insulating phase against on-site Coulomb repulsions.

(3) At 3/4-filling, a pair of sites connected by a $t_+$-bond forms a spin-1/2 isospin state, namely,
$\frac{1}{\sqrt{2}}(|\uparrow\downarrow, \uparrow\rangle+|\uparrow,\uparrow\downarrow\rangle)$ or
$\frac{1}{\sqrt{2}}(|\uparrow\downarrow, \downarrow\rangle+|\downarrow,\uparrow\downarrow\rangle)$,
which yields a double degeneracy in the ES for  $t_+$-bonds, even in the insulating phase.
These isospins then form an antiferromagnetic Mott insulating state through couplings mediated by $t_-$-bonds.
Although the relationship between the bulk excitation spectrum and the ES is non-trivial 
and generally lacks a strict one-to-one correspondence~\cite{Alba12,Chandran14}, 
the low-lying ES can qualitatively reflect the underlying bulk physical phases under specific configurations.
This behavior of isospins forming a Mott insulator is highly reminiscent of the half-filled Mott insulator in the ordinary Fermi-Hubbard model~\cite{Cha18}, 
leading to a closely resembling ES structure characterized by a gapped charge sector but a gapless spin sector.

(4) For $t_-$-bonds at 1/2-filling, the largest Schmidt eigenvalue, $\lambda_1$, is close to 1, implying that the system is nearly represented
by a product state of two subsystem states separated by the $t_-$-bond, with no diverging correlation length.
As a result, the gap between  $\lambda_1$ and the second largest eigenvalue, $\lambda_2$, is quite large.
Additionally, we observe an almost quadruple degeneracy for the $\lambda_2$ level.
This structure can be understood through a perturbative framework where the weak $t_-$-bond across the virtual boundary cut acts as a perturbation~\cite{Alba12}. 
In the zeroth-order limit, the decoupled subsystems directly yield the single dominant eigenvalue $\lambda_1$. 
In the first-order correction, a single particle-hole pair is created across the partition line between the nearest neighboring sites. 
There are four independent configurations to form this pair, as a particle with either spin-up or spin-down can hop across the boundary 
from left to right or from right to left. 
This spin and directional degeneracy naturally gives rise to the observed almost quadruple degeneracy in $\lambda_2$. 
This feature strongly reflects the robustness of the band gap inherited from the non-interacting SSH model, 
indicating that the low-lying ES faithfully mirrors single-electron or single-hole excitations 
rather than collective, strongly correlated modes.
All these features, which are quite different from those observed in the Mott insulating phase at 3/4-filling,
indicate that at 1/2-filling, the system forms a gapped band insulating phase. 

\begin{figure}[b] 
\includegraphics[width=0.45\textwidth]{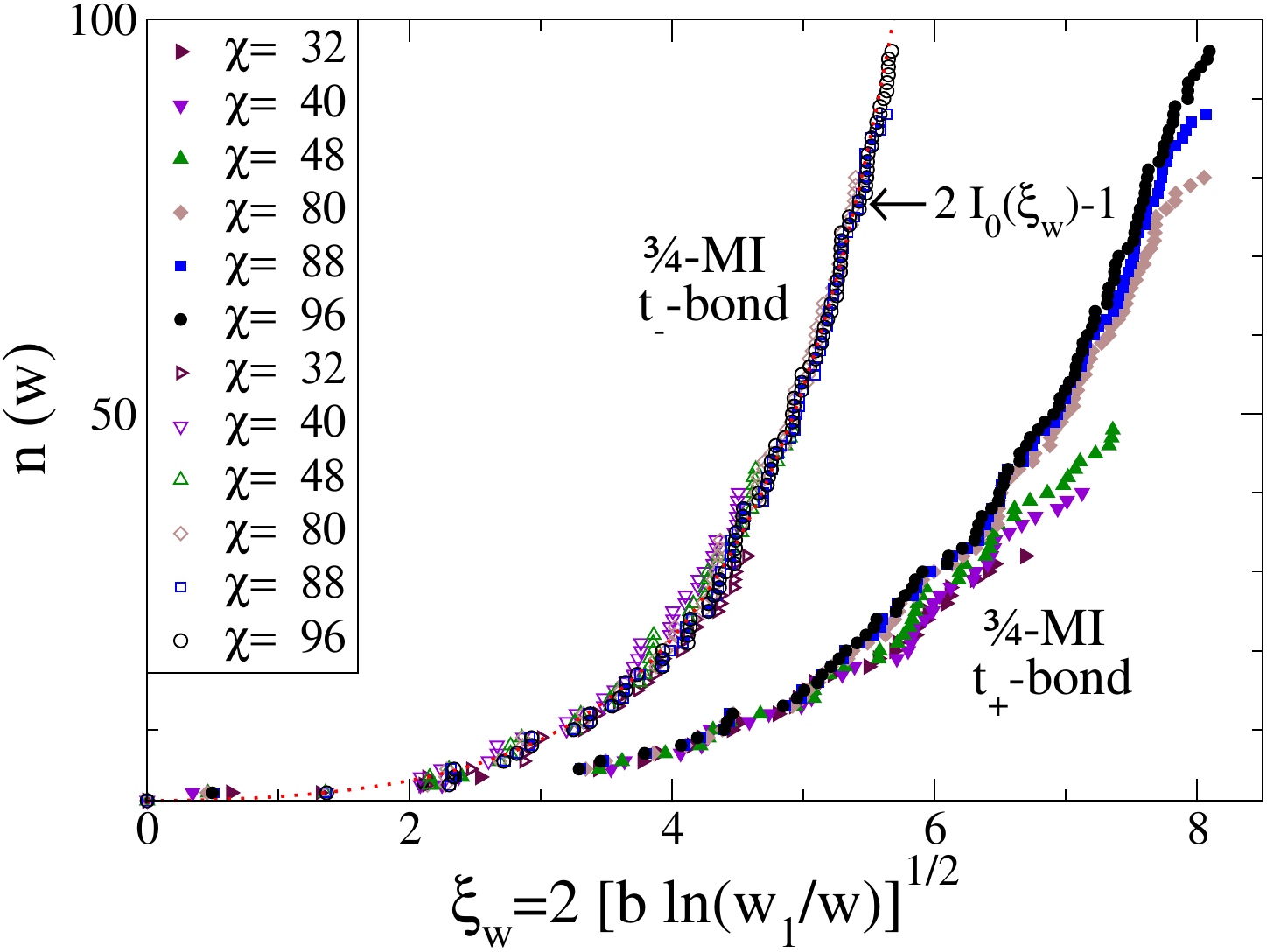}
\caption{The behavior of $n(w)$, the mean number of eigenvalues larger than or equal to a given $w$, for different $\chi$,
showing a universal scaling behavior for $t_-$-bonds in the 3/4-filled Mott insulating phase.
No universal behavior is observed for $t_+$-bonds and in other phases (not shown). 
The dotted line represents the theoretical prediction given by $2I_0(\xi_w)-1$ with a global degeneracy parameter of $g=2$.}
\label{fig:1dFHM_Sn0}
\end{figure}

Above, we have discussed that the insulating state at 3/4-filling is a Mott insulator in which the spin-1/2 isospins are weakly coupled through $t_-$-bonds. 
Therefore, we expect the half-chain ES to exhibit the scaling behavior theoretically predicted by conformal field theory~\cite{Calabrese08,Alba18}.
Indeed, Figure~\ref{fig:1dFHM_Sn0} shows this scaling behavior for $t_-$-bonds in the Mott insulating phase at 3/4-filling.
However, we do not observe this scaling behavior for the $t_+$-bonds (i.e., the data for different $\chi$ do not collapse onto a single curve).
This scaling behavior is also absent at 1/2-filling (not shown).
In the figure, $n(w)$ represents the mean number of eigenvalues of the reduced density matrix (with $w_\alpha\equiv \lambda_\alpha^2$)
that are larger than or equal to a given $w$ for different values of $\chi$, specifically comparing the $t_-$-bond (open symbols) and the $t_+$-bond (closed symbols) at 3/4-filling.
The dotted line represents the theoretical prediction given by $2I_0 (\xi_w)-1$
for one-dimensional critical systems with a global degeneracy parameter $g=2$,
where $I_0(\xi_w)$ is the modified Bessel function of the first kind,
$\xi_w \equiv 2 \sqrt{b \ln(w_1/w)}$, and $b \equiv -\ln(w_1)$.

The scaling properties of the ES exhibit the following two features.
First, the ES for $t_-$-bond at 3/4-filling shows a universal distribution. 
This confirms that the system at this filling behaves as a spin-1/2 antiferromagnetic Mott insulator, 
in sharp contrast to the half-filled band insulator that lacks such universal scaling properties.
Second, the universal behavior at 3/4-filling exhibits a global degeneracy parameter of $g=2$. 
Note that this 'global degeneracy'~\cite{Alba18} does not imply a strict two-fold degeneracy of individual eigenvalues, 
but rather characterizes the bulk scaling behavior of the cumulative number of eigenvalues $n(w)$.

The excellent agreement with the universal scaling function further supports the physical picture 
that the low-energy physics of these effective degrees of freedom closely mimics that of spin-1/2 particles 
in a non-dimerized half-filled antiferromagnetic Mott insulator
where $g=2$.
The emergence of $g=2$ in our dimerized model can be qualitatively interpreted through a multi-channel scaling behavior. 
In this regime, the lattice modulation is expected to modify the effective Luttinger parameters
and decoupling channels in the spin and charge sectors, 
as conceptually suggested in related interacting SSH-like frameworks~\cite{Jin23, He21}. 
A comprehensive analytical investigation via full bosonization remains an intriguing topic for future research.

\section{Summary}
We study the entanglement properties of the one-dimensional dimerized Fermi-Hubbard model through MPS representations.
A phase diagram is constructed by investigating the EE,
revealing two insulating phases at 1/2-filling and 3/4-filling, separated by an intervening metallic phase.
These phases are conclusively identified by their distinct structures in the ES:
the metallic phase exhibits a double degeneracy in its low-lying levels,
whereas in the insulating phases, this degeneracy is selectively lifted depending on the bond configurations due to the opening of a bulk charge gap.
Although the EE provides an intuitive guide for the edge modes in the fully dimerized limit, 
the definitive topological features are anchored by the ES. 

Furthermore, the finite-entanglement scaling behavior of the half-chain EE 
reveals the fundamentally different nature of the two insulating phases:
the insulating phase at 3/4-filling is a Mott phase exhibiting a characteristic logarithmic scaling, 
whereas that at 1/2-filling is a band insulating phase characterized by a short and finite correlation length. 
This distinction is further confirmed by the universal distribution of the ES observed at 3/4-filling, 
which exhibits a global degeneracy parameter of $g=2$.

All these results demonstrate that entanglement properties, 
when synthesized through both macroscopic entropy scaling and microscopic spectral degeneracies, 
provide a powerful framework for characterizing the quantum and topological phases of interacting systems.

\begin{acknowledgments}
This work was supported by the Basic Science Research
Program through the National Research Foundation of Korea
funded by the Ministry of Education, Science and Technology (Grant No. NRF-2019R1F1A1062704 to MCC,
Grant No. NRF-2021R1F1A1056081 to JWL,
and Grant No. NRF-2021R1F1A1052347 to MHC).
\end{acknowledgments}


\end{document}